\documentclass[%
aip,
amsmath,amssymb,
reprint,%
]{revtex4-1}

\usepackage{graphicx}
\usepackage{dcolumn}
\usepackage{bm}

\usepackage[utf8]{inputenc}
\usepackage[T1]{fontenc}
\usepackage{mathptmx}
\usepackage{etoolbox}

\makeatletter
\def\@email#1#2{%
	\endgroup
	\patchcmd{\titleblock@produce}
	{\frontmatter@RRAPformat}
	{\frontmatter@RRAPformat{\produce@RRAP{*#1\href{mailto:#2}{#2}}}\frontmatter@RRAPformat}
	{}{}
}%
\makeatother
\begin{document}
	
	\preprint{AIP/123-QED}
	
	\title[]{Wideband Brillouin light scattering analysis of spin waves excited by a white-noise RF generator}
	\author{Lukáš Flajšman}
	\email{lukas.flajsman@aalto.fi}
	\affiliation{ 
		NanoSpin, Department of Applied Physics, Aalto University School of Science, P.O. Box 15100, FI-00076 Aalto,
		Finland
	}%
	\author{Ondřej Wojewoda}%
	\affiliation{ 
		CEITEC BUT, Brno University of Technology, 612 00 Brno, Czech Republic
	}%
	
	\author{Huajun Qin}
	\affiliation{ 
		School of Physics and Technology, Wuhan University, Wuhan 430072, China	
		Wuhan Institute of Quantum Technology, Wuhan 430206, China
	}%
	\affiliation{ 
		Wuhan Institute of Quantum Technology, Wuhan 430206, China
	}%
	\author{Kristýna Davídková}%
	
	\affiliation{ 
		Institute of Physical Engineering, Brno University of Technology, 616 69 Brno, Czech Republic
	}%
	
	
	\author{Michal Urbánek}%
	\affiliation{ 
		CEITEC BUT, Brno University of Technology, 612 00 Brno, Czech Republic
	}%
	\affiliation{ 
		Institute of Physical Engineering, Brno University of Technology, 616 69 Brno, Czech Republic
	}%
	\author{Sebastiaan van Dijken}
	\affiliation{ 
		NanoSpin, Department of Applied Physics, Aalto University School of Science, P.O. Box 15100, FI-00076 Aalto,
		Finland
	}%
	
	\date{\today}
	
	\begin{abstract}
		Spin waves are studied intensively for their intriguing properties and potential use in future technology platforms for the transfer and processing of information and microwave signals. The development of devices and materials for spin-wave systems requires a lot of measurement time and effort, and thus increasing the measurement throughput by extending the instrumentation capabilities is of the essence. In this letter, we report on a new and straightforward approach to increase the measurement throughput by fully exploiting the wideband detection nature of the Brillouin light scattering technique in single-shot experiments using a white-noise RF generator.   
	\end{abstract}
	
	\maketitle
	
	The Brillouin light scattering (BLS) technique has been used for decades as a tool to characterize the properties of spin waves or magnons (quanta of spin waves) in magnetic materials\cite{Hillebrands1999,Sebastian2015}. In micro-BLS\cite{Sebastian2015}, inelastic scattering of photons on spin waves provides information on the spin-wave momentum, frequency and phase with a spatial resolution of approximately 300 nm. Implementation of the BLS technique relies on precise spectrometers (scanning Fabry-Pérot\cite{Sebastian2015} or more recently virtually imaged phase array (VIPA) spectrometers\cite{Yan2020}) for the detection of minute changes in the scattered-light frequency with high contrast. Because BLS works in the frequency domain, it records the entire spin-wave spectrum in a one-stage pass of the spectrometer (Fabry-Pérot type) or a single image (VIPA type). BLS stands out from other spin-wave measurement techniques by its ability to detect non-coherent waves originating, e.g., from thermal\cite{Sandercock1978, Hillebrands1987, Cochran1988} or spin-orbit-torque\cite{Divinskiy2018} excitation. Thermal excitation offers a truly wideband and flat spin-wave spectrum. The probing of thermal spin waves by BLS therefore allows for the recording of spin-wave spectra in single-shot experiments. Unfortunately, the low signal level\cite{Hache2020a} and non-directional excitation of thermal spin waves are incompatible with studies on the properties of propagating spin waves in complex geometries such as waveguides\cite{Demidov2008, Demidov2009,Wojewoda2020, Turcan2021}, bends\cite{Sadovnikov2017} and resonators\cite{Qin2021}. The intriguing frequency properties of these magnonic systems can be fully characterized by external excitation. This is conventionally done by connecting a RF generator to a microwave antenna fabricated on top of the sample and performing sweeps of successive single-frequency measurements\cite{Wojewoda2020,Turcan2021,Sadovnikov2017}. The complete spectral response of the system is obtained by changing the RF frequency in a step-by-step fashion. For detailed spin-wave measurements, this calls for PC-based automation, resulting in delays originating from synchronization issues and the non-optimal use of the spectrometer as it scans over an extended frequency range while spin waves are excited at a fixed frequency. The accumulation of these effects results in long measurement times. 
	
	\begin{figure}
		\includegraphics{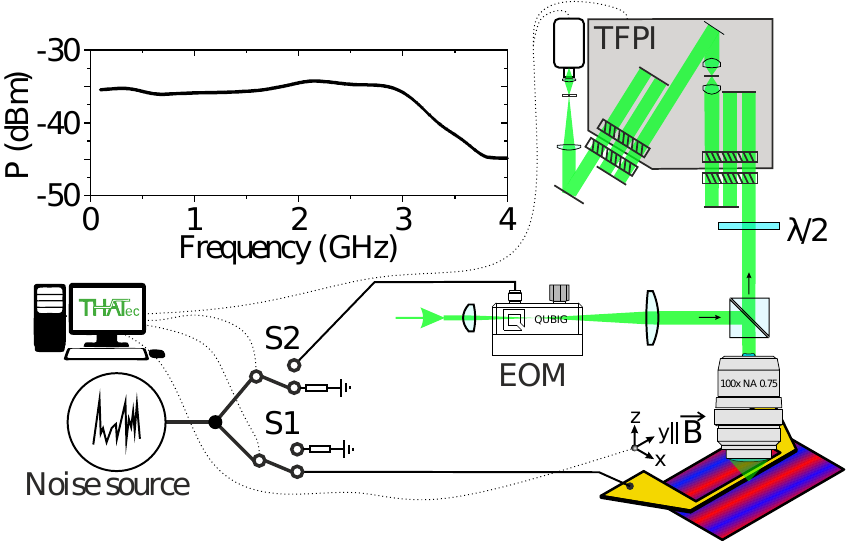}
		\caption{\label{Fig1}  Simplified scheme of the experimental setup with combined optical and RF branches. The laser light goes through the electro-optical phase modulator (EOM) placed in the beam expander. After passing a beam splitter the light is focused on the sample by a microscope objective. The back-reflected scattered light passes the half-wave plate (mounted in a rotational mount) before it enters the tandem Fabry-Pérot interferometer (TFPI). Here the fine spectrum is analyzed with high contrast. The RF signal from the noise source is split into two arms. Each arm contains a single microwave switch (S1 and S2). One branch connects to the antenna fabricated on the sample, and the other feeds to the EOM. A measurement of the total RF power provided by the noise source is shown in the top-left corner. }
	\end{figure}
	
	Here, we introduce a new approach for measuring spin-wave properties in single-shot BLS experiments. Integral to our method is the replacement of the RF generator by a white-noise RF source (Pasternack PE85N1012, see schematic in Fig.~\ref{Fig1}). In combination with a microstrip antenna, the white-noise RF source provides wideband and directional excitation of propagating spin waves, thus combining the advantages of thermal and RF excitation while avoiding their drawbacks. We demonstrate the use of a white-noise RF generator in a commercial phase-resolved micro-focused BLS system (THATec Innovation GmbH). Although the setup and measurement process are greatly simplified by replacing the RF generator with a white-noise source, we show that even advanced studies including phase-resolved BLS\cite{Serga2006,Vogt2009} can still be performed. Because we excite and detect all frequencies within the noise-source excitation range and BLS detection window, we obtain large amounts of spin-wave data in single-shot measurements. The limiting factor of our approach is given by the frequency resolution of the spectrometer. 
	
	We test our approach by measuring the excitation and propagation of spin waves in a continuous liquid-phase-epitaxy (LPE) grown\cite{Dubs2017} 100-nm-thick yttrium iron garnet (YIG) film on a gadolinium gallium garnet (GGG) substrate. The spin waves are excited by a 500-nm-wide microwave antenna with a 5 nm Ta/55 nm Au structure. The antenna is patterned by electron beam lithography and lift-off and connects to the white-noise source via a RF cable and RF probe.
	
\begin{figure}
		\includegraphics{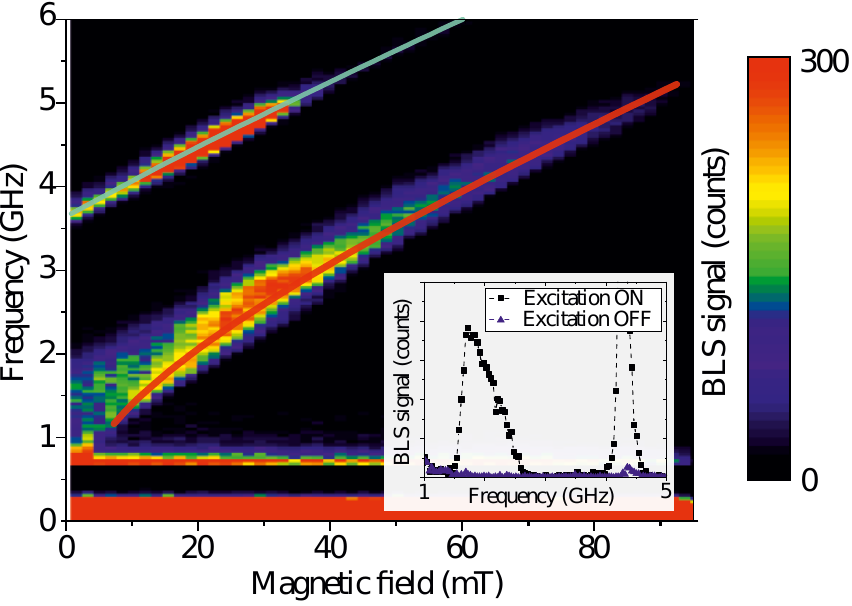}
		\caption{\label{Fig2} BLS intensity map of excited spin-wave modes as a function of external magnetic field. Two main spin-wave branches are clearly visible. The upper branch is the first perpendicular standing spin wave mode and the lower branch is the Damon-Eshbach mode. The lines are Levenberg-Marquardt fits of the Herring-Kittel formula\cite{Qin2018} to the signal maxima found in the individual branches. The inset shows single BLS spectra obtained at 10 mT field with the white-noise RF generator on (black squares) and off (blue triangles).}
	\end{figure}

	In the first study, we positioned the BLS laser spot close to the center of the microwave antenna and swept the magnetic field (oriented along the microwave antenna) from 0 mT to 95 mT. For each value of the external field we recorded the BLS spectrum for two conditions: with the white-noise source off (S1 and S2 open) and with the white-noise source on (S1 closed, S2 open). All other parameters in the experiment were kept constant. The BLS intensity map recorded with the white-noise source on is shown in Fig.~\ref{Fig2}.
	
	The BLS map shows two clear spin-wave branches. The bottom branch represents propagating Damon-Eshbach (DE) spin waves in the YIG film. The onset of this branch corresponds to the ferromagnetic resonance frequency (FMR). The upper branch is the first perpendicular standing spin-wave (PSSW) mode. Comparing the BLS signal recorded with and without noise (see inset of Fig.~\ref{Fig2} and supplementary Fig.~\ref{FigS1}), we observe a well-resolved signal with high signal-to-noise ratio for active spin-wave excitation by the noise source and almost no signal when the noise source is disconnected (only thermal spin-wave excitation). Because YIG is almost transparent at the BLS laser wavelength (532 nm), the back-scattering cross-section of thermal spin waves is too small to be resolved by BLS\cite{Hache2020a}. 
	
	To extract magnetic parameters, we fit the frequency position of the maximum BLS signal of the two spin-wave branches [$f\left(\mu_0 H_{\mathrm{ext}}\right)$] using the Herring-Kittel formula\cite{Qin2018}. Fits for the propagating spin waves and the PSSW mode are performed simultaneously to decorrelate the joint effect that some parameters have on the two branches. For fixed parameter $\gamma/2\pi=28\, \mathrm{GHz/T}$, this analysis yields $M_{\mathrm{S}}=\left(148\pm 2\right)\, \mathrm{kA/m},\, A_{\mathrm{ex}}=\left(3.64\pm 0.1 \right)\, \mathrm{pJ/m},\,t=\left(96\pm 3 \right)\, \mathrm{nm}$ and $B_0=\left(0.7\pm 0.3 \right)\, \mathrm{mT}$ for the saturation magnetization, exchange constant, film thickness and effective anisotropy field. Information on the spin-wave dispersion can be obtained from the shape of the lower spin-wave band. However, this requires hard-to-find information on the spin-wave density of states, the excitation power factor of the antenna, and the BLS sensitivity factor\cite{Demokritov2006a, Wojewoda2022}. Later we will demonstrate a straightforward extraction of the spin-wave dispersion from single-shot phase-resolved BLS measurements.
	
\begin{figure}
		\includegraphics{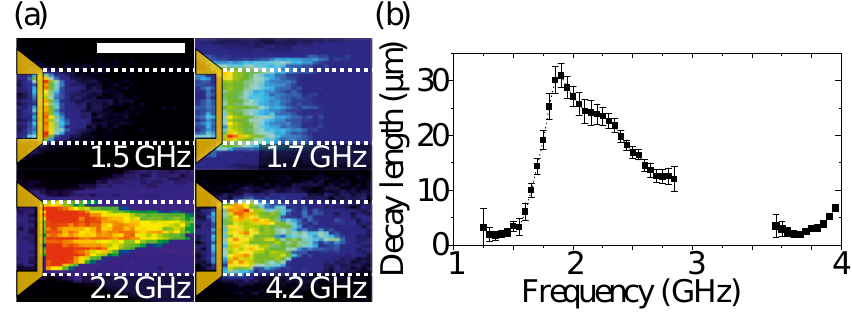}
		\caption{\label{Fig3} (a) Spatial BLS intensity maps recorded at 17 mT external field for four excitation frequencies. The microwave antenna is schematically overlaid on the maps. The scale bar in the top left panel indicates 5~µm. The dotted lines mark the area from which data are extracted for further analysis. (b) Extracted spin-wave decay length as a function of frequency.}
	\end{figure}

	To further demonstrate the capabilities of BLS with a white-noise RF generator, we recorded spatial BLS intensity maps capturing the propagation of spin waves [Fig.~\ref{Fig3}(a)] excited by a short and narrow microwave antenna. In the experiments, we scanned the sample under the BLS laser spot in 300 nm steps while keeping the white-noise source on (S1 closed, S2 open). For each position, the BLS spectrum from 0 GHz to 4.5 GHz is recorded in a single shot. Since we excite and detect all modes simultaneously, it is possible to reconstruct (demultiplex) individual maps of propagating spin waves with a frequency step that is limited only by the setting of the spectrometer (50 MHz in our case). Selected exemplary maps are shown in Fig.~\ref{Fig3}(a). For a full set of spatial BLS intensity maps at different frequency, see supplementary Fig.~\ref{FigS2}.
	
	The BLS maps show clear intensity signals that decay away from the microwave antenna. Well below the FMR frequency ($\sim$1.9 GHz), only direct excitation in the vicinity of the microwave antenna is measured [see BLS map for 1.5~GHz in Fig.~\ref{Fig3}(a)]. Starting at 1.7 GHz, caustic-like modes are emitted from the corners of the antenna (Fig.~\ref{Fig3}(a)). Above the FMR frequency, the antenna also excites propagating DE spin waves. Because the microwave antenna is short, interference of the caustic beams and the DE mode produces a pronounced focusing of the spin-wave intensity [see BLS map for 2.2~GHz in Fig.~\ref{Fig3}(a)], as previously observed in other works\cite{Vanatka2021,Bertelli2020,Shiota2020}. With increasing frequency, the focal point of spin-wave intensity moves closer to the microwave antenna. The 500-nm-wide antenna does not excite spin waves with a wavelength below $\approx$700 nm, leading to an abrupt drop in BLS intensity above 2.7 GHz (see supplementary Fig.~\ref{FigS2}). A BLS signal originating from the first PSSW mode is measured above 4 GHz [see BLS map for 4.2~GHz in Fig.~\ref{Fig3}(a)].
	
	Wideband data collection during a single BLS scan can be used for high-throughput extraction of spin-wave parameters. As an example, we determine the frequency dependence of the spin-wave decay length. We do this by line-averaging the BLS signal in the rectangular area marked by white dashed lines in Fig.~\ref{Fig3}(a) and fitting the resulting line profiles to an exponentially decaying spin-wave intensity\cite{Heinz2019}:
	\begin{eqnarray}
		I_{\mathrm{SW}}=I_{\mathrm{i}}\exp \left(\frac{2x}{\lambda_{\mathrm{D}}}\right)+I_0
		\label{eq:one}.
	\end{eqnarray}
	Here, $I_{\mathrm{i}}$ is the initial spin-wave intensity, $I_{\mathrm{0}}$ is the background intensity originating from detector noise and thermal spin waves, and $\lambda_{\mathrm{D}}$ is the spin-wave decay length. Figure~\ref{Fig3}(b) summarizes the extracted spin-wave decay length. The initial increase of the decay length below the FMR frequency represents data for the caustic-like beams that are emitted from the corners of the microwave antenna. The maximum decay length is obtained near the FMR frequency. The subsequent decrease of the decay length is caused by a reduction of the Damon-Eshbach spin-wave group velocity with increasing frequency, but also by stronger focusing of the spin-wave intensity at high frequency. Due to the latter effect, the extracted spin-wave decay length is smaller than in other works utilizing longer antenna structures on similar YIG films\cite{Yu2015, Qin2018}.
	
	
	The unique potential of the BLS technique does not end with the spatial mapping of the spin-wave intensity. It has been shown before that the BLS method can be easily extended to the visualization of the spin-wave phase\cite{Sebastian2015, Serga2006,Vogt2009}. This is done by connecting an electro-optic modulator to the same RF source. The EOM produces sidebands to the laser spectrum with a constant phase. Because the sidebands stemming from propagating spin waves depend on the wave phase, interference of the two signals can be exploited to extract the phase information. This concept, which is well understood for spin-wave excitation by single-tone RF generators, can be extended to our BLS technique with a white-noise RF source. For this to work, the optical signals from the EOM and sample must be mutually coherent. Similar to white-light interferometry, we achieved coherency in our setup by closely matching the path lengths of the two RF signal paths. When this condition is met, phase-resolved spin-wave measurements can be conducted over a broad frequency range in a single-shot experiment. To demonstrate this type of measurement, we first roughly matched the intensity of the spin-wave signal (S1 closed, S2 open in Fig.~\ref{Fig1}) and the EOM signal (S1 open, S2 closed) by rotating the $\lambda/2$ plate. After this, we activated both switches and scanned the sample over the same area as in Fig.~\ref{Fig3}. An example of a phase-resolved BLS map is shown in Fig.~\ref{Fig4}(a) (for a full set of images see supplementary Fig.~\ref{FigS3}). The phase-resolved map of propagating spin waves clearly shows the fringes produced by the interference of the EOM and BLS signals.
	
	\begin{figure}
		\includegraphics{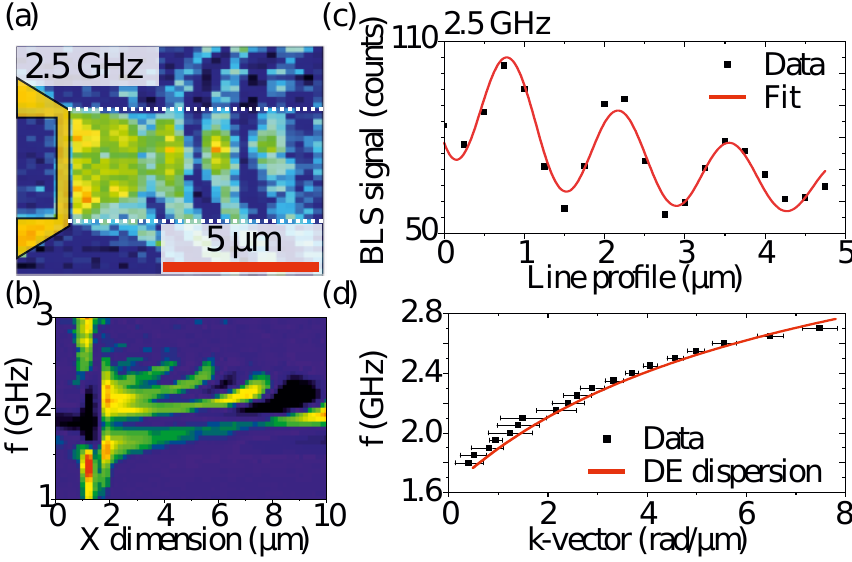}
		\caption{\label{Fig4} (a) Phase-resolved BLS map recorded at 17 mT external field and 2.5 GHz. The microwave antenna is illustrated in yellow and the white dotted lines mark the area used for data extraction. (b) Map of spin-wave line profiles as a function of frequency. (c) Spin-wave line profile at 2.5 GHz and fit to the data. (d) Measured spin-wave dispersion and the calculated dispersion of DE spin waves in the YIG film.}
	\end{figure}
	
	The wideband BLS technique introduced here captures detailed phase information on the complex interference pattern between the caustic-like beams and the DE spin waves above the FMR frequency. Besides focusing of the spin-wave intensity, the interference of these modes also produces a significant bending of the spin-wave wavefront [Fig.~\ref{Fig4}(a)]. Despite the use of a short antenna causing such effects, we now demonstrate that single-scan BLS data collection with a white-noise RF generator enables fast measurements of the spin-wave dispersion.  Figure~\ref{Fig4}(b) shows the frequency dependence of the line-averaged phase-resolved BLS signal, which we derived from BLS maps like the one depicted in Fig.~\ref{Fig4}(a) by averaging the signal over the area marked by the dashed white lines. Focusing on the excitation of DE spin waves at frequencies ranging from 1.9 GHz to 3.0 GHz, we clearly observe a decreasing wavelength with increasing frequency. The data in Fig.~\ref{Fig4}(b) also reveal a well-defined phase relation with an almost constant initial offset at all frequencies. We fit the spin-wave profiles of individual frequency bins by a two-wave interference model:
	\begin{eqnarray}
		I=I_{\mathrm{SW}}+I_{\mathrm{EOM}}+2\sqrt{I_{\mathrm{EOM}}I_{\mathrm{SW}}} \cos\left(2\pi \frac{x}{\lambda}+\theta_0\right),
		\label{eq:two}
	\end{eqnarray}
	where $\lambda$ is the spin-wave wavelength and $\theta_0$ represents the initial phase offset between the EOM and the spin-wave signal. As an example, we show an experimental line profile measured at 2.5 GHz and a model fit in Fig.~\ref{Fig4}(c). The derived wave vector ($k=2\pi/\lambda$) is plotted for different excitation frequencies in Fig.~\ref{Fig4}(d) together with an analytical calculation of the DE spin-wave dispersion in the YIG film based on the magnetic parameters extracted previously ($M_{\mathrm{S}},\, A_{\mathrm{ex}},\,t$ and $B_0$)\cite{Flajsman2020, Kalinikos1986}. The good fit to the experimental data without any tuning of the magnetic parameters demonstrates the power of our high-throughput BLS measurement technique.
	

	
	In summary, we present a new twist to the well-established field of BLS-based spin-wave characterization. By replacing the single-tone RF generator by a relatively cheap wideband white-noise RF source we reduce the measurement time by exciting and detecting full spectra of spin waves in a single BLS scan. We demonstrate the capabilities of the method by analyzing the field dependence of spin-wave modes and recording BLS maps of the spin-wave intensity and phase. The frequency resolution of the method, which is limited by the spectrometer, allows for single-shot measurements of the spin-wave dispersion relation. With these promising characteristics, we are confident that BLS setups with an integrated white-noise RF source will become a popular technique for high-throughput measurements of spin-wave properties.
	
	\begin{acknowledgments}
		This work was supported by the Academy of Finland (Grant No. 338748). We acknowledge CzechNanoLab Research Infrastructure supported by MEYS CR (LM2018110). O.W. was supported by Brno PhD talent scholarship and acknowledge support from the project Quality Internal Grants of BUT (KInG  BUT), Reg. No. CZ.02.2.69/0.0/0.0/19073/0016948, which is financed from the OP RDE and EF19\_073/0016948. Authors greatly appreciate the provision of the YIG samples by Carsten Dubs, INNOVENT e.V. Technologieentwicklung Jena.
	\end{acknowledgments}
	
	\section*{Data Availability Statement}
	
	The data that support the findings of this study are available from the corresponding author upon reasonable request.
	\bibliography{library2}
	
	\appendix
	
	\section{Supplementary information}
	
	\renewcommand{\thefigure}{S\arabic{figure}}
	\setcounter{figure}{0} 
	
	\begin{figure*}
		\includegraphics{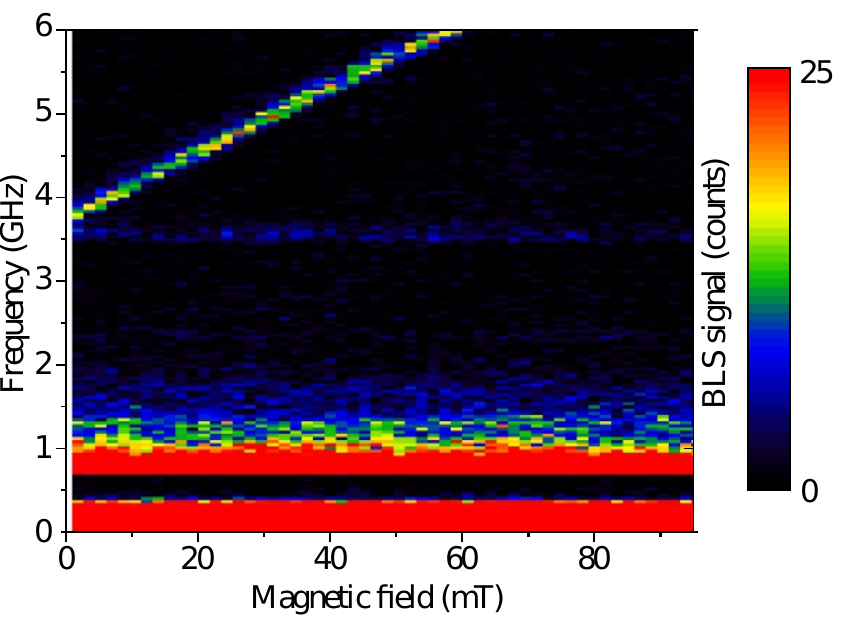}
		\caption{\label{FigS1} BLS intensity map of excited spin-wave modes as a function of magnetic bias field without external excitation (only thermal spin waves are excited).}
	\end{figure*}
	\begin{figure*}
		\includegraphics[width=.9\textwidth]{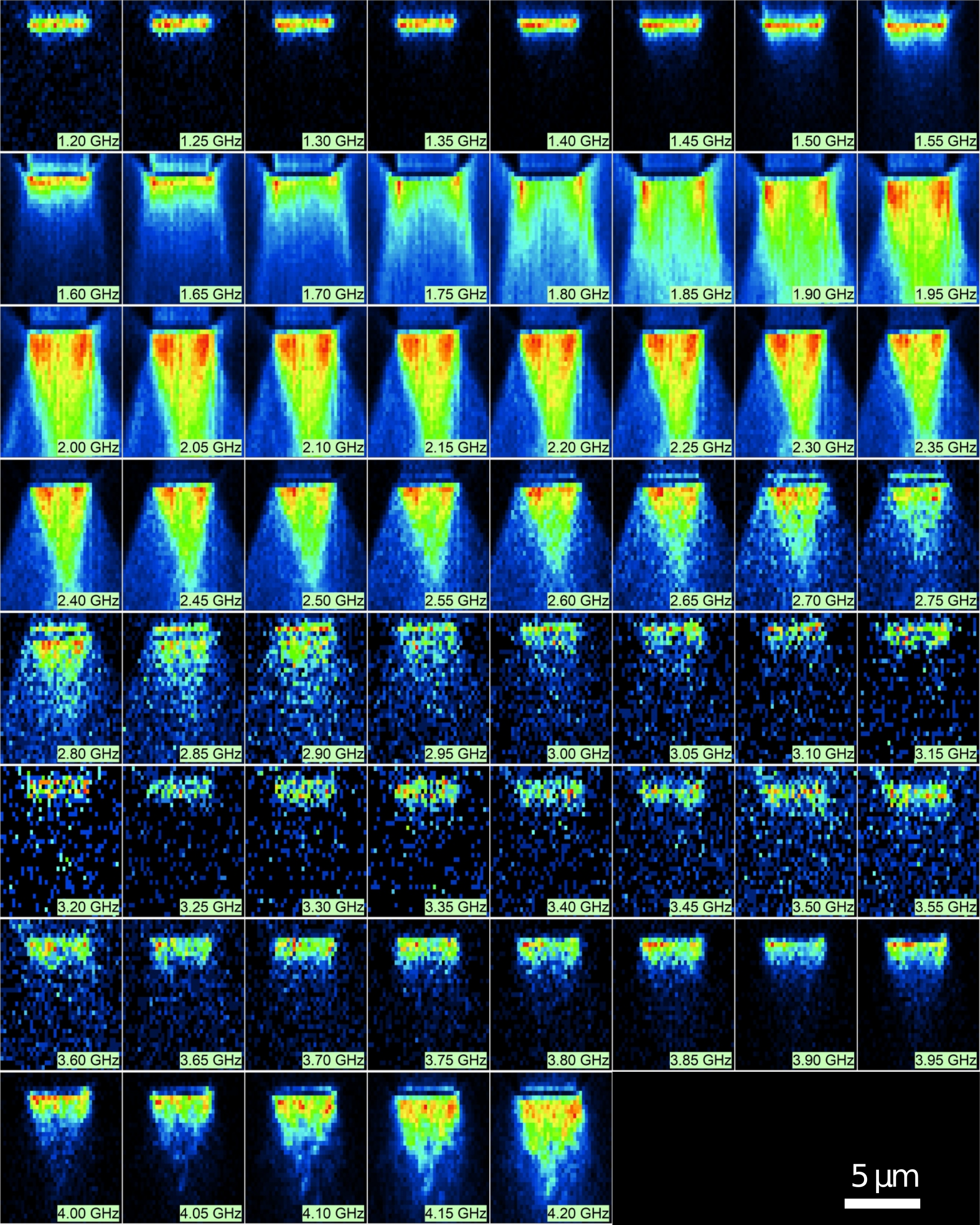}
		\caption{\label{FigS2}Full set of BLS intensity maps recorded at 17 mT external field with the white-noise RF generator on. The color scale of individual maps is normalized to the highest signal value. The BLS signal measured at low frequency ($\sim$1.20 GHz $-$ 1.55 GHz) is caused by non-resonant direct excitation of the YIG film. Upon approaching the FMR frequency ($\sim$1.60 GHz $-$ 1.90 GHz), caustic-like beams are emitted from the antenna corners. Above the FMR frequency ($\sim$1.90 GHz $-$ 2.70 GHz), we measure propagating DE spin waves. Interference of this mode and the caustic-like beams from the antenna corners leads to a focusing of the spin-wave intensity. The signal above $\sim$4 GHz originates from the first PSSW mode.}
	\end{figure*}
	\begin{figure*}
		\includegraphics[width=.9\textwidth]{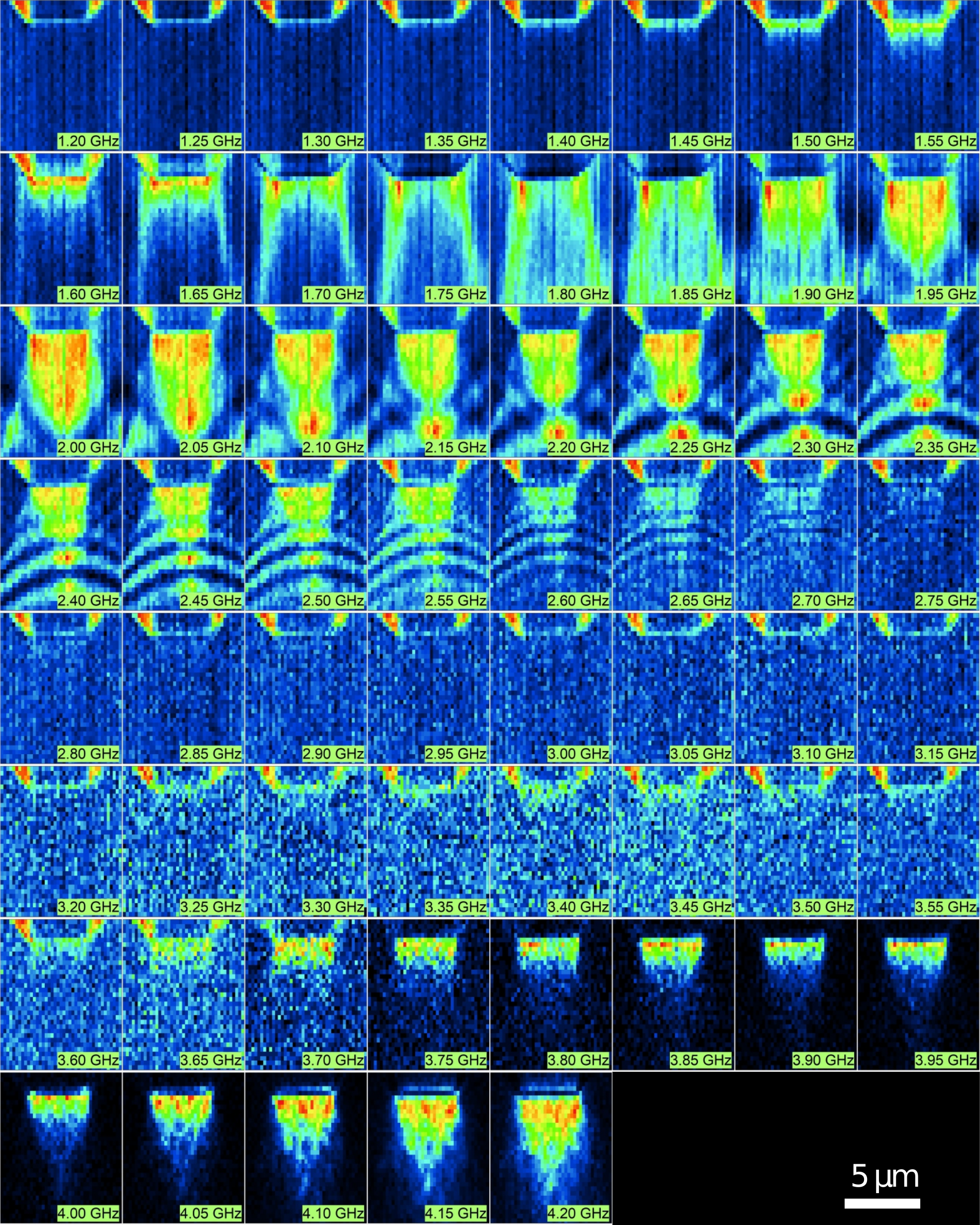}
		\caption{\label{FigS3} Full set of phase-resolved BLS maps recorded at 17 mT external field with both the white-noise RF generator and the EOM on. The color scale of the individual maps is normalized to the highest signal value. The signal measured at low frequency ($\sim$1.20 GHz $-$ 1.55 GHz) is caused by the reflection of the EOM signal from the microwave antenna. Complex interference patterns with curved wavefronts form above $\sim$1.90 GHz when the caustic-like beams and propagating DE spin waves are excited. The wavelength of the propagating mode decreases with increasing frequency. The signal from the first PSSW mode above $\sim$4 GHz does not show interference fringes.}
	\end{figure*}
	

\end{document}